\documentclass[prb,twocolumn,superscriptaddress,longbibliography]{revtex4-2}
\usepackage{amsmath}
\usepackage{amssymb}
\usepackage{amsthm}
\usepackage{amsfonts}
\usepackage{listings}
\usepackage{enumerate}
\usepackage{latexsym}
\usepackage{psfrag}
\usepackage{bm} 
\usepackage[all]{xy}
\usepackage{subfigure}
\usepackage{color}
\usepackage{mathtools}
\usepackage{array}
\usepackage{placeins}
\usepackage{hyperref}

\usepackage{physics}
\usepackage{tikz}
\usepackage{adjustbox}
\usetikzlibrary{calc}
\usetikzlibrary{decorations.pathreplacing}

\begin{document}

\title{Gauge symmetry of excited states in projected entangled-pair state 
simulations}

\author{Yi Tan}
\affiliation{Department of Physics, Southern University of Science and Technology, Shenzhen 518055, China}

\author{Ji-Yao Chen}
\email{chenjiy3@mail.sysu.edu.cn}
\affiliation{Center for Neutron Science and Technology, Guangdong Provincial Key Laboratory of Magnetoelectric Physics and Devices, School of Physics, Sun Yat-sen University, Guangzhou 510275, China}

\author{Didier Poilblanc}
\email{didier.poilblanc@irsamc.ups-tlse.fr}
\affiliation{Laboratoire de Physique Th\'eorique, C.N.R.S. and Universit\'e de Toulouse, 31062 Toulouse, France}

\author{Jia-Wei Mei}
\email{meijw@sustech.edu.cn}
\affiliation{Department of Physics, Southern University of Science and Technology, Shenzhen 518055, China} 
\affiliation{Shenzhen Key Laboratory of Advanced Quantum Functional Materials and Devices, Southern University of Science and Technology, Shenzhen 518055, China}
\date{\today}

\begin{abstract}
While gauge symmetry is a well-established requirement for representing topological orders in projected entangled-pair state (PEPS), its impact on the properties of low-lying excited states remains relatively unexplored. Here we perform PEPS simulations of low-energy dynamics in the Kitaev honeycomb model, which supports fractionalized gauge flux (vison) excitations. We identify gauge symmetry emerging upon optimizing a PEPS ground state. Using the PEPS adapted local mode approximation, we further classify the low-lying excited states by discerning different vison sectors. Our simulations of spin and spin-dimer dynamical correlations establish close connections with experimental observations. Notably, the selection rule imposed by the locally conserved visons results in nearly flat dispersions in momentum space for 
excited states belonging to the 2-vison or 4-vison sectors.
\end{abstract}

\maketitle

\emph{Introduction --}
The fundamental concept of long-range entanglement forms the cornerstone of our exploration into topological orders within quantum phases of matter~\cite{Kitaev2006,Levin2006,Wen2019}. The Kitaev spin liquid~\cite{Kitaev2006a} provides a fertile ground for unraveling topological properties, including topological degeneracy and fractionalized excitations, with numerous potential material realizations~\cite{Jackeli2009,Chaloupka2010,Takagi2019}. The allure of topological excitations extends into the realm of spin dynamics~\cite{Baskaran2007,Knolle2014a,Knolle2014}, prompting investigations through inelastic neutron scattering (INS) and light (Raman and resonant-inelastic-X-ray scattering (RIXS)) measurements on Kitaev materials~\cite{Banerjee2016,Banerjee2017,Sandilands2015,Nasu2016,Glamazda2016,Li2019,Pei2020,Kim2020,Torre2023}. Thus, simulations of low-lying excitations in the topologically ordered phases are highly valuable for establishing a close connection with experimental observations.

Projected entangled-pair state (PEPS) provides a robust framework for capturing the intricate topology and quantum entanglement in two-dimensional many-body systems~\cite{Verstraete2004, Cirac2021}. The additive negative topological entanglement entropy of topologically ordered states~\cite{Kitaev2006,Levin2007} is a manifestation of the gauge symmetry constraint inherent in PEPS~\cite{Schuch2010,Cirac2021}, allowing the construction of the degenerate ground state manifold~\cite{Chen2010,Schuch2010,Schuch2012,Jiang2015,Mei2017,Lee2019,Chen2020,Francuz2020} and associated boundary Hamiltonian~\cite{Schuch2013} and the computation of modular matrices~\cite{He2014,Mei2017,Crone2020}. Recent advancements have expanded the applicability of PEPS to excited states~\cite{Vanderstraeten2015,Vanderstraeten2019a,Ponsioen2020,Ponsioen2022,Ponsioen2023b,Tu2023a} through incorporating the single-mode approximation~\cite{Feynman1953a}. In this framework, the ground state is locally perturbed by an ``impurity'' tensor, and momentum states, generated by superpositions of these local perturbations, provide natural representations of low-energy excited states~\cite{Chi2022}. While achieving a faithful representation of gauge symmetry in PEPS is feasible through variational optimization~\cite{Crone2020,Hasik2022}, the exploration of gauge structure and, consequently, fractionalized excitations in low-energy excited states remains relatively uncharted.

Here, we delve into the intricate gauge structure of PEPS simulations, with a particular emphasis on unraveling fractionalized excitations within the Kitaev honeycomb model. This model is characterized by a gauge symmetry involving flux (vison) operators and hosts a vortex-free spin liquid ground state~\cite{Kitaev2006a}. The optimized ground state PEPS maintains its essential vortex-free nature, on top of which, we construct low-energy excited states by introducing local impurity tensors and label them with the associated vison number. While our simulations of spin dynamics align commendably with exact solutions, the spin-dimer dynamics simulations offer valuable insights for  Raman and RIXS experiments. 

\begin{table*}[htbp]
    \caption{\label{tab:table1}
    Variational ground state energy per site $E_0$ (in units of $J$) and the expectation of the flux operator $w_p=\langle \hat{W}_p\rangle$.}
    \begin{ruledtabular}
        \begin{tabular}{cccccccc}
        & $Exact$ & $D = 2$ & $D=4$ & $D = 2$ & $D = 3$ & $D = 4$ & $D = 6$  \\
        & & (loop-gas)& (loop-gas) & (optimized)& (optimized)& (optimized) & (optimized) \\ \colrule
        $E_0$ & -0.78730 & -0.65399 & -0.78576& -0.75624 & -0.77194 & -0.78695 & -0.78716 \\
        $w_p $ & 1  & 1 & 1 & 0.02533 & 0.88148 & 0.999996 & 0.99997\\        
        \end{tabular}
    \end{ruledtabular}
\end{table*}

\emph{Gauge symmetry in ground state PEPS--}
The Kitaev honeycomb model~\cite{Kitaev2006a} is defined as
\begin{eqnarray}
  \label{eq:KHM}
  H = -J \sum_{\langle ij\rangle_\gamma} \sigma_i^\gamma \sigma_j^\gamma,
\end{eqnarray}
where $\sigma^\gamma$ represents the Pauli matrices with $\gamma = x, y, z$ for the nearest-neighbbor bonds $\langle ij \rangle_\gamma$, along the $x$, $y$ and $z$ directions. In this work, we focus on the ferromagnetic coupling with $J>0$. On the honeycomb lattice, a variational ground state can be represented in the PEPS form as $|\Psi(A)\rangle = \sum_{s_1,\cdots,s_N}{\rm tTr} A_{abc}^{s_i}|s_1\cdots s_i\cdots s_N\rangle$, and further graphically expressed as
\begin{eqnarray}
\label{eq:GPEPS}    
\includegraphics{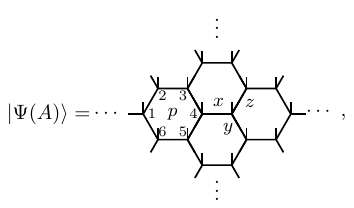},
\end{eqnarray}
where each local tensor $A_{abc}^{s}$ comprises one physical spin-1/2 index ($s$) and three virtual indices ($a$, $b$, and $c$) with bond dimension $D$. The PEPS virtual legs are along the nearest-neighbor $x$, $y$, and $z$ bonds as depicted in Eq.(\ref{eq:GPEPS}). The Kitaev model is distinguished by the gauge symmetry characterized by flux operators $\hat{W}_p = \sigma_1^x\sigma_2^y\sigma_3^z\sigma_4^x\sigma_5^y\sigma_6^z$ linked to the $p$-th hexagon plaquette in Eq.(\ref{eq:GPEPS}). Due to $\hat{W}_p^2=1$, their eigenvalues are $w_p=\pm1$ with $\mathbb{Z}_2$ gauge nature. For $w_p=+1$, the plaquette is vortex-free, while for $w_p=-1$, it possesses a vortex (vison). Under periodic boundary conditions, the product of all flux operators equals 1, i.e., $\prod_p \hat{W}_p=1$. This topological constraint imposes limitations, allowing only excitations with an even vison number, indicating a $\mathbb{Z}_2$ gauge symmetry in the local tensor to encode the topological property~\cite{Schuch2010}.

\begin{figure}[b]
    \centering
    \includegraphics[width = \columnwidth]{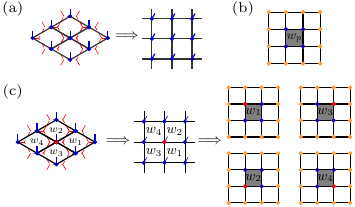}
    \caption{(a) Mapping the honeycomb lattice to the square lattice through the combination of the A-B sublattices. CTMRG evaluation of the expectation value $w_p$ of the vortex operator for the ground state PEPS in (b), and excited-state PEPS in (c). }
    \label{wp_diag}
\end{figure}

The optimization of the ground state PEPS is accomplished through variational energy minimization techniques, employing gradient optimization methods~\cite{Vanderstraeten2016,Corboz2016}. Here, we use the corner transfer matrix renormalization group (CTMRG) method~\cite{Nishino1996,Orus2009,Corboz2014} for tensor contractions with a truncation dimension $\chi=64$. The energy gradient is obtained through reverse mode automatic differentiation~\cite{Liao2019}. In contrast to prior work~\cite{Lukin2023}, where optimization proceed directly in the honeycomb lattice, our approach employs a more general framework by mapping the system to square lattice, as shown in Fig.~\ref{wp_diag}(a). 
Table.~\ref{tab:table1}  lists the results of the ground-state variational energy and the flux operator expectation. Additionally, results from the loop-gas PEPS wave function, featuring a meticulously designed gauge structure~\cite{Lee2019}, are included for comparison. For a small bond dimension $D=2$, the optimized PEPS exhibits a lower variational energy than the loop-gas PEPS. However, this comes at the cost of losing gauge symmetry, as evidenced by the flux expectation $w_p$, which is evaluated as in Fig.~\ref{wp_diag}(b), being close to zero and deviating from the exact result $w_p=1$. With increasing bond dimension $D$, the variational energy is further optimized as anticipated, and concurrently, we observe the flux expectation $w_p$ approaching the exact result $w_p=1$ for $D=4$, thereby displaying the essential vortex-free nature of the Kitaev spin liquid ground state. Further improvement for the energy can be achieved with $D=6$, while the flux expectation remains essentially unchanged.

\emph{Gauge symmetry in excited state PEPS-- }
From the ground state, local excited state PEPSs are constructed as $|B_i\rangle=\sum_{abca'b';s_{\alpha_i},s_{\beta_i}}B_{aba'b'}^{s_{\alpha_i},s_{\beta_i}}\frac{\partial}{\partial (A_{abc}^{s_{\alpha_i}}A_{a'b'c}^{s_{\beta_i}})}|\Psi(A)\rangle$ as illustrated in the graph
\begin{eqnarray}
  \label{eq:peps_es}
  \includegraphics{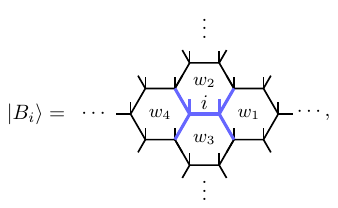}
\end{eqnarray}
where the impurity tensor $B$ acts on the two sites $\alpha_i$ and $\beta_i$ within the $i$-th unit cell. 

\begin{figure}[b]
    \centering
    \includegraphics[width = \columnwidth]{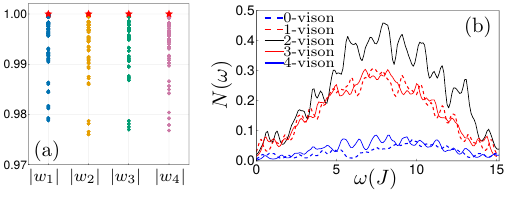}
    \caption{
    (a) Absolute value distributions of $w_{1,2,3,4}$, with red stars denoting the results for the ground state PEPS with $D=4$.
    (b) Zero momentum low-energy density of state with a broadening $\sigma = 0.2J$ for the 0, 1, 2, 3, and 4-vison sectors. The full density of state is expected to be captured up to an energy of order 5-7 roughly, above which other excitations not included here are expected to contribute.}
    \label{fig:wp_dos}
\end{figure}   

\begin{figure*}[htbp]
    \centering
    \includegraphics[width = 2\columnwidth]{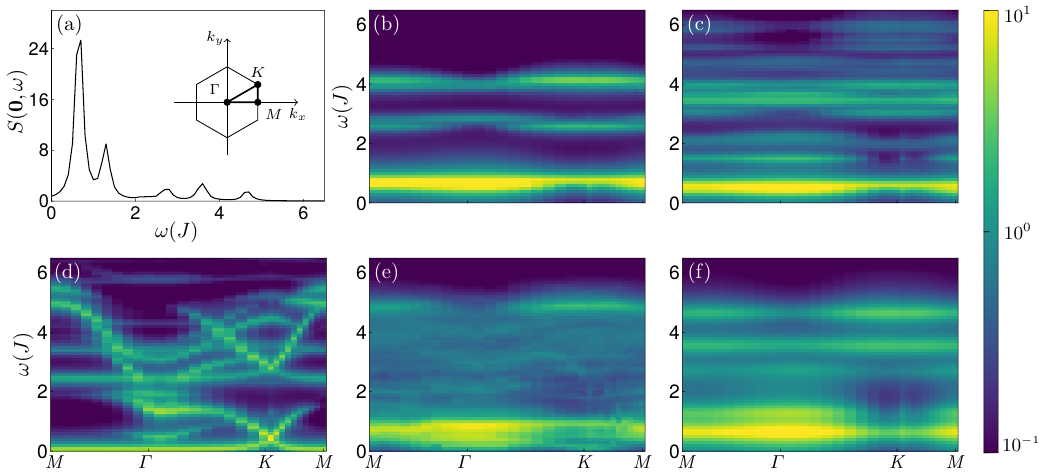}
    \caption{Computed spin dynamics with a Lorentzian broadening $\sigma = 0.1J$. (a) Zero momentum dynamics for optimized PEPS with $D=4$.  The first Brillouin zone and the momentum path for the spectral function are shown in the inset. Momentum-resolved spin dynamics for loop-gas PEPS with $D=2$ in (b) and $D=4$ in (c), optimized  PEPS with $D=2$ in (d), $D=3$ in (e), and $D=4$ in (f).}
\label{fig:all_spectra}
\end{figure*}

Once the ground state is the vortex-free state $|\psi(A);{w_p=1}\rangle$, the excited state in Eq. (\ref{eq:peps_es}) can be characterized by the eigenvalues $w_{1,2,3,4}$ of the four plaquette operators $\hat{W}_{1,2,3,4}$, denoted as $|B_i;\{w_{1,2,3,4}\}\rangle$. Other plaquettes are not perturbed, and thus $w_p=1$ for $p\notin\{1,2,3,4\}$, which follows from the symmetry in the local tensor (see discussions in Appendix.~\ref{sec:loop_gas}). The eigen-equations $\hat{W}_p|B_i\rangle= w_p|B_i\rangle$ for $p=1,2,3,4$ can be diagonalized simultaneously as the flux operators commute with each other. 
In the CTMRG contraction, the honeycomb lattice is transformed into a square lattice through the combination of the A-B sublattices in Fig.~\ref{wp_diag}(a), 
and the expectation $w_p$ of the vortex operator for the ground state is evaluated in Fig.~\ref{wp_diag}(b). For the excited states, we evaluate $w_{1,2,3,4}$ in Eq.(\ref{eq:peps_es}) by placing the impurity tensor on the four adjacent positions, respectively, as the Fig.~\ref{wp_diag}(c).

For the infinite PEPS simulation, our choice of initial boundary conditions in the CTMRG allows both even and odd vison sectors (see Appendix~\ref{sec:boundary_condition} for details).
$w_{1,2,3,4}$ take the value of $\pm1$, independent of each other, and $|B_i;\{w_{1,2,3,4}\}\rangle$ has $2^4=16$ different vison configurations. Fig.~\ref{fig:wp_dos}(a) illustrates the absolute value distributions of $w_{1,2,3,4}$ for the optimized PEPS with $D=4$, in which all values closely approximate the exact value of 1 within a numerical error of less than 2.5\%, affirming the gauge symmetry in both ground and excited PEPSs. Note that a small violation of the gauge symmetry in the optimized PEPS could yield a small deviation of exact commutativity of the $\hat{W}_p$ and of the projectivity $\hat{W}_p^2=1$, leading to the small deviations in $w_p$.

By taking a linear superposition of all locally perturbed states, we can construct translational-invariant excited states $|\mathfrak{B}_{\mathbf{q}}\rangle = \sum_i e^{i\mathbf{q}\cdot\mathbf{r}_i}|B_i\rangle$ with momentum $\mathbf{q}$.  
In the variational space spanned by these excited states (hereafter labeled by $m$), the generalized eigenvalue equation for local tensor $B_i^m$ is given by
\begin{eqnarray}
\label{eq:ham}
\mathcal{H}_\mathbf{q} B^m_{i} =\lambda_{\mathbf{q}} \mathcal{N}_{\mathbf{q}} B^m_i,
\end{eqnarray}
where $\mathcal{H}_{\mathbf{q}}$ and $\mathcal{N}_{\mathbf{q}}$ represent the effective Hamiltonian and norm matrices.

The locally conserved vison makes the Hamiltonian (\ref{eq:ham}) block-diagonal. 
For better clarity, we categorize the 16 vison configurations into distinct vison sectors, spanning 0 to 4 visons. Each sector corresponds to a specific number of vison configurations: 1 for 0-vison, 4 for 1-vison, 6 for 2-vison, 4 for 3-vison, and 1 for 4-vison. The total density of state (at momentum $\mathbf{q}=\mathbf{0}$) shown in Fig.\ref{fig:wp_dos}(b) within each sector reveals fractionalized continua, indicative of multiple-spinon excitations.

\begin{figure*}[htbp]
    \centering
    \includegraphics[width = 2\columnwidth]{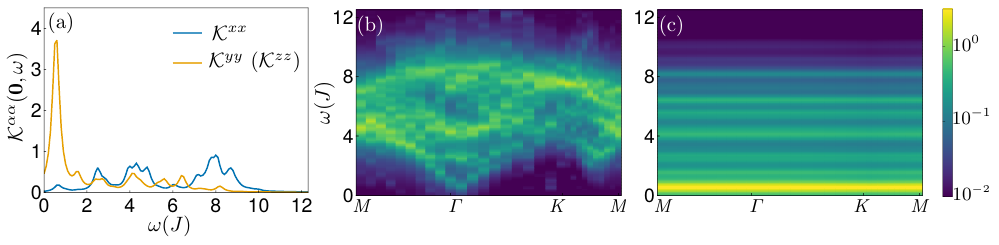}
    \caption{Computed spin dimer dynamics with a Lorentzian broadening $\sigma = 0.2J$ for an optimized PEPS with $D=4$; (a) $\mathcal{K}^{xx}(\mathbf{0}, \omega)$ and $\mathcal{K}^{yy}(\mathbf{0}, \omega)$ ($\mathcal{K}^{zz}(\mathbf{0}, \omega)$);   
    (b) color plot of $\mathcal{K}^{xx}(\mathbf{q}, \omega)$; (c) color plot of $\mathcal{K}^{yy}(\mathbf{q}, \omega)$ ($\mathcal{K}^{zz}(\mathbf{q}, \omega)$). }
\label{fig:dimer_spectra}
\end{figure*}
\emph{Spin dynamics --} Spin excited states $\sigma_s^\gamma|\psi(A)\rangle$ on the $s$-site can be also labeled by the surrounding vison configurations $w_{1,2,3,4}$ since the flux operator $\hat{W}_p$ commutes or anti-commutes with local operators.
Examining the plaquettes in Eq.(\ref{eq:GPEPS}),
when $\sigma^x$ is applied on the site 2, 3, 5, or 6, we have $\sigma_s^x \hat{W}_p\sigma_s^x=-\hat{W}_p$; when $\sigma^x$ acts on the site 1 or 4, we have $\sigma_s^x \hat{W}_p\sigma_s^x=\hat{W}_p$.
Consequently, the spin excited state $\sigma_s^x|\psi(A)\rangle$ introduces a vison in each of the two plaquettes containing the bond along the $x$-direction, adjacent to the $s$-site~\cite{Baskaran2007,Knolle2014a}. Similar outcomes apply for $\sigma_s^y|\psi(A)\rangle$ and $\sigma_s^z|\psi(A)\rangle$, leading to the following
\begin{eqnarray}
\label{eq:spinv}
    (w_1,w_2,w_3,w_4)=\left\{ \begin{aligned}
    (+1,-1,-1,+1), & \quad \sigma_{\alpha_i}^x|\psi(A)\rangle\\
    (+1,-1,-1,+1), & \quad \sigma_{\beta_i}^x|\psi(A)\rangle\\
    (+1,-1,+1,-1), & \quad \sigma_{\alpha_i}^y|\psi(A)\rangle\\
    (-1,+1,-1,+1), & \quad \sigma_{\beta_i}^y|\psi(A)\rangle\\
    (+1,+1,-1,-1), &\quad \sigma_{\alpha_i}^z|\psi(A)\rangle\\
    (-1,-1,+1,+1), &\quad \sigma_{\beta_i}^z|\psi(A)\rangle \end{aligned}\right. .    
\end{eqnarray}
The dynamical spin correlation function is defined as
\begin{eqnarray}
\label{eq:spin}
    S^{\gamma\gamma}_{ij}(t)=\langle \psi(A)|\sigma_j^\gamma(t)\sigma_i^\gamma|\psi(A)\rangle,
\end{eqnarray}
from which we obtain the spin dynamics
\begin{eqnarray}
    S(\mathbf{q},\omega)=\sum_{m\gamma}|\langle\mathfrak{B}^m_\mathbf{q}|S_{\mathbf q}^\gamma\rangle|^2\delta(\omega-(\lambda_{\mathbf{q}}^m-\lambda_0)).
\end{eqnarray}
Figure~\ref{fig:all_spectra} presents the numerical results on the spin dynamics $S(\mathbf{q},\omega)$ for different ground state PEPSs. Fig.~\ref{fig:all_spectra}(a) displays $S(0,\omega)$ for the optimized PEPS with $D=4$ in which the peak around $\omega = 0.5 J$ corresponds to the 2-vison-related excitations.

The spin excited states belong to the 2-vison sector. Vison excitations, static and conserved over time, adhere to the gauge symmetry, leading to an ultra-short spin correlation with a strictly vanishing of correlations beyond nearest neighbors~\cite{Baskaran2007,Knolle2014a}. This ultra-short correlation manifests in flat momentum-resolved spin dynamics, providing a crucial validation for our numerical findings. Faithfully capturing the gauge structure in both the ground-state and excited PEPS, as exemplified by the analytical loop-gas wave function in Fig. \ref{fig:all_spectra} (b) and (c) and the optimized wave function with $D=4$ in Fig. \ref{fig:all_spectra}(f), the observed flat pattern in spin dynamics signifies an exceptionally weak dependence on momentum. In contrast, the gauge structure is entirely lost in the optimized wave function with $D=2$, resulting in dispersive spin dynamics in Fig.~\ref{fig:all_spectra} (d). It is noteworthy that despite the loop-gas wave function with $D=2$ providing a less accurate variational ground state, it accurately captures the gauge structure, ensuring a correct representation of short spin correlation physics in Fig.~\ref{fig:all_spectra} (b)
(see Appendix~\ref{sec:loop_gas} for further details for the gauge structure of excited state based on the loop-gas PEPS.)

\emph{Spin-dimer dynamics --} From Eq.(\ref{eq:spinv}), we can easily show the following for the dimer excited states
\begin{eqnarray}
\label{eq:dimerv}
    (w_1,w_2,w_3,w_4)=\left\{ \begin{aligned}(+1,+1,+1,+1), & ~ \mathcal{K}^x_{i}|\psi(A)\rangle\\
    (-1,-1,-1,-1), &~ \mathcal{K}^y_{i} |\psi(A)\rangle\\
    (-1,-1,-1,-1), &~ \mathcal{K}^z_{i}|\psi(A)\rangle \end{aligned}\right. ,    
\end{eqnarray}
with $\mathcal{K}^\gamma_{i}=\sigma_{\alpha_i}^\gamma\sigma_{\beta_i}^\gamma$ and $i$ refers to an x-bond. 
Therefore, $\mathcal{K}_i^x|\psi(A)\rangle$ belongs to the 0-vison sector, whereas $\mathcal{K}_i^y|\psi(A)\rangle$ and $\mathcal{K}_i^z|\psi(A)\rangle$ reside in the 4-vison sector~\cite{Knolle2014}. These distinct vison sectors result in markedly different spin-dimer correlations which are defined as
\begin{eqnarray}
    \mathcal{K}^{\gamma\gamma}_{ij}(t)=\langle \psi(A)|\mathcal{K}^\gamma_{j}(t)\mathcal{K}^\gamma_{i}|\psi(A)\rangle,
\end{eqnarray}
from which we have the spin-dimer dynamical structural factor
\begin{eqnarray}
\label{eq:KR}
    \mathcal{K}^{\gamma\gamma}(\mathbf{q},\omega)=\sum_{m}|\langle\mathfrak{B}^m_{\mathbf{q}}|\mathcal{K}^\gamma_{\mathbf{q}}\rangle|^2\delta(\omega-(\lambda_{\mathbf{q}}^m-\lambda_0)).
\end{eqnarray}

Figure~\ref{fig:dimer_spectra} presents the spin-dimer dynamics for the optimized PEPS with $D=4$. The zero-momentum spin-dimer dynamics $\mathcal{K}(0,\omega)$, which is potentially probed in the Raman scattering measurements~\cite{Sandilands2015,Nasu2016,Glamazda2016,Li2019,Pei2020}, displays a 4-vison peak around $0.5J$ in $\mathcal{K}^{yy}$ and $\mathcal{K}^{zz}$, as shown in Fig.~\ref{fig:dimer_spectra}(a). The momentum-dependent spin-dimer dynamics, potentially measurable in RXIS experiments~\cite{Kim2020,Torre2023}, are depicted in Fig.~\ref{fig:dimer_spectra}(b)-(c). 
Remarkably, the spin-dimer dynamics $\mathcal{K}^{xx}(\mathbf{q},\omega)$ depicted in Fig.\ref{fig:dimer_spectra}(b) displays a dispersive fractional continuum in the zero-vison sector spanning the entire Brillouin zone.
This characteristic feature is in agreement with the model Eq.~\eqref{eq:KHM} being gapless~\cite{Kitaev2006a}, and provides a distinctive signature of spinon-pair excitations, particularly relevant for interpreting RIXS measurements.
In contrast, the selection rule imposed by vison conservation -- all excitations created by $\mathcal{K}_i^y$ and $\mathcal{K}_i^z$ belong to the 4-vison sector -- results in a flat momentum-resolved dynamics ($\mathcal{K}^{yy}(\mathbf{q},\omega)$ and $\mathcal{K}^{zz}(\mathbf{q},\omega)$), 
as seen in Fig.\ref{fig:dimer_spectra}(c)\footnote{For Raman and RIXS experiments, $\mathcal{K}^{xx}(\mathbf{q},\omega)$, $\mathcal{K}^{yy}(\mathbf{q},\omega)$ and $\mathcal{K}^{zz}(\mathbf{q},\omega)$ are simultaneously detected and not easily disentangled. Thus we expect the experimental spectra to reveal the flat dispersions embedded in the fractionalized continuum. }. The distinctive momentum dependences of the spin-dimer dynamics highlight the primordial role of the gauge structure in the optimized PEPS with $D=4$.

\emph{Conclusion --}
In conclusion, our investigation has delved into the intricate gauge structure of projected entangled pair state (PEPS) simulations, extending our scrutiny beyond the ground state to 
low-lying excited states within the Kitaev honeycomb model. Importantly, we have demonstrated that gauge symmetry is not maintained solely in the optimized ground state achieved through variational optimization; rather, it seamlessly extends to excited states constructed by the PEPS adaptive local mode approximation method. 
Furthermore, our simulations of low-energy dynamics in spin and spin-dimer correlations establish connections with experimental observations such as INS and light scattering (Raman and RIXS) experiments. The selection rule imposed by locally conserved vison flux yields flat momentum-resolved dynamics if the excited states contain vison excitations. 

This study underscores the efficacy of employing PEPS to explore gauge symmetry and fractionalized excitations within topologically ordered states, presenting a robust framework for further investigations in the realm of quantum many-body systems. For general topologically ordered states, although there is no associated local conserved quantity the same as vortex operator in the Kitaev spin liquid, we can still identify the emergent internal gauge symmetry. Utilizing the gauge flux generated by this symmetry, analogous to the role of $\hat{W_p}$ in the Kitaev model, we can extend our method to explore gauge symmetries of the excited states in these topological phases.

\begin{acknowledgments}
This work is supported by the National Key Research and Development Program of China (Grant No. 2021YFA1400400), Shenzhen Fundamental Research Program (Grant No. JCYJ20220818100405013), the Guangdong Innovative and Entrepreneurial Research Team Program (Grants No. 2017ZT07C062), Shenzhen Key Laboratory of Advanced Quantum Functional Materials and Devices (Grant No. ZDSYS20190902092905285), Guangdong Basic and Applied Basic Research Foundation (Grant No. 2020B1515120100). 
J.-Y.C. was supported by Open Research Fund Program of the State Key Laboratory of Low-Dimensional Quantum Physics (project No.~KF202207), Fundamental Research Funds for the Central Universities, Sun Yat-sen University (project No.~23qnpy60), a startup fund from Sun Yat-sen University, the Innovation Program for Quantum Science and Technology 2021ZD0302100, Guangzhou Basic and Applied Basic Research Foundation (grant No.~2024A04J4264), and National Natural Science Foundation of China (NSFC) (grant No.~12304186). 
Part of the calculations reported were performed on resources provided by the Guangdong Provincial Key Laboratory of Magnetoelectric Physics and Devices, No.~2022B1212010008.
This work was also supported by the TNTOP ANR-18-CE30-0026-01 grant awarded by the French Research Council.
\end{acknowledgments}

\appendix

\section{Vison excitation under different boundary conditions}
\label{sec:boundary_condition}

\begin{figure}[htbp]
    \centering
    \includegraphics[width = 0.95\columnwidth]{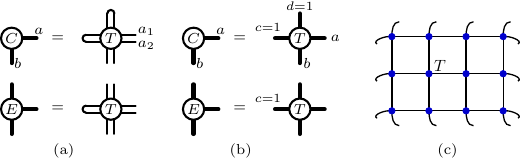}
    \caption{Tensor network evaluation under the different boundary conditions. (a-b) Open boundary condition in CTMRG. The initial boundary tensors are chosen as: transfer matrices of the trace of the redundant bond ($\sum_x T_{xxab}$) in (a), and transfer matrices of the first index of the redundant bond ($T_{11ab}$) in (b). (c) Periodical boundary condition in a $3\times4$ torus, where each tensor includes a unit cell of the honeycomb lattice.
    }
\label{fig:boundary}
\end{figure}

The vison excitations depend on the boundary conditions through the selection of initial boundary tensors, as shown in Fig.~\ref{fig:boundary}(a-b) in our infinite PEPS simulation. Odd-vison excitations are only allowed when the initial boundary tensors are chosen as depicted in Fig.~\ref{fig:boundary}(a), while they are not permitted in Fig.~\ref{fig:boundary}(b). In the latter case, the odd-vison will have exactly zero norm, thereby prohibiting the existence of odd-vison states. Additionally, for periodic boundary conditions, our exact contraction evaluation on a $3\times4$ torus in Fig.~\ref{fig:boundary}(c) indicates existence only of even-vison states. The allowed odd-vison excitations in the boundary conditions of Fig.~\ref{fig:boundary}(a) is likely due to vison condensation on the boundary\cite{kitaev2012models}, which needs further investigations.

\section{Gauge structure of excited states for the loop-gas PEPS}
\label{sec:loop_gas}

In this section, we interpret our result in terms of the analytical loop-gas PEPS for the Kitaev spin liquid~\cite{Lee2019}, in which the gauge symmetry is explicitly incorporated. 

In an effort to preserve the symmetries of the Kitaev spin liquid ground state, a straightforward naive trial wave function for the Kitaev model is the GHZ state along the (111)-direction 
\begin{eqnarray}
    |{\rm GHZ}\rangle=\bigotimes_i |(111)\rangle_i+\bigotimes_i |(-1-1-1)\rangle_i.
\end{eqnarray}
When we rotate the (111)-direction into the $\tilde{z}$-axis in the rotated axis, where the spin indices are rotated as $|\tilde{s}\rangle=\mathcal{S}|s\rangle$ with 
 $\mathcal{S}=\exp(-i\frac{\pi}{8}\sigma^z)\exp(-i\frac{\theta}{2}\sigma^y)$  ($\cos\theta=\frac{1}{\sqrt{3}}$),
 it becomes straightforward to write down the non-zero elements of the local tensor for the GHZ state $T_{111}^{\tilde{s}=\uparrow}=T_{222}^{\tilde{s}=\downarrow}(2,2,2)=1$. We obtain the local tensor for the (111)-GHZ as $T_{abc}^s=\sum_{\tilde{s}}\mathcal{S}_{s\tilde{s}}T_{abc}^{\tilde{s}}$ by applying the spin-rotation operator  $\mathcal{S}$.

The (111)-GHZ state not only exhibits inadequate variational energy but also lacks the crucial property of being an eigenstate of the vortex operator $W_p$, a characteristics essential for maintaining the vortex-free nature of the Kitaev spin-liquid ground state.  
To address the vortex-related challenge, Lee et al. introduced the loop gas operator $\hat{Q}_{\rm LG}$ layer atop the GHZ state with the local tensor
\begin{eqnarray}
    \includegraphics{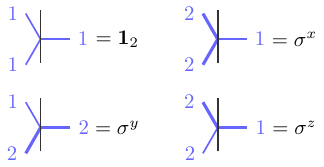} \quad ,
\end{eqnarray}
which preserves the symmetries of the Kitaev model, and additionally possesses the $\mathbb{Z}_2$ gauge structure
\begin{gather}
    \includegraphics{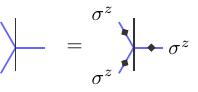} ~, \\
    \includegraphics{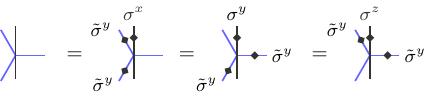} ~, 
\end{gather}
with $\tilde{\sigma}^y=(\sigma^x-\sigma^y)/\sqrt{2}$ and satisfying the following eigen equations
\begin{eqnarray}
\label{eq:gz}
    &&\includegraphics{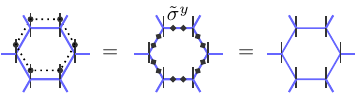}, \nonumber\\
    &&\includegraphics{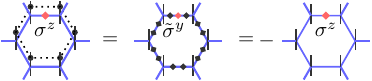},
\end{eqnarray}
where \adjustbox{valign=c}{\includegraphics{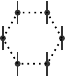}} denotes the vortex operator $W_p$.
The loop gas operator $\hat{Q}_{\rm LG}$ enriches the gauge symmetry of the GHZ state. When applied to the GHZ state, it yields the gauged GHZ state as the loop-gas PEPS
\begin{eqnarray}
  |{\rm{GHZ}}\rangle_{\mathfrak{g}} = \hat{Q}_{\rm LG}|\rm{GHZ}\rangle,  
\end{eqnarray}
 with the local tensor \adjustbox{valign=c}{\includegraphics{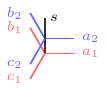}}, which can be readily
 demonstrated to manifest a vortex-free character according to the eigen
 equations in Eq.(\ref{eq:gz}).
 
Numerically verifying the ground state of the loop-gas PEPS, we confirm its vortex-free nature with $w_p=1$. Additionally, we examine the excited states with local impurity tensors, finding that they exhibit $w_p=\pm1$ for $p=1,2,3,4$. Specifically, for the $D=2$ loop-gas PEPS, where the GHZ state is replaced by the product state $\bigotimes_i |(111)\rangle_i$, our numerical check involves both CTMRG contraction on an infinite lattice and exact contraction on a small $3\times4$ unit cluster. In the case of the torus-shaped cluster with periodic boundary conditions, excited states in the odd (1 and 3) vison sectors display zero norm. Conversely, for the cluster with open boundary conditions, excited states in the odd (1 and 3) vison sectors exhibit non-zero norm.

The local impurity tensors for the spin-excited states are
\begin{eqnarray}
\label{eq:spinEs}
    |S_i^x\rangle&=& ~\adjustbox{valign=c}{\includegraphics{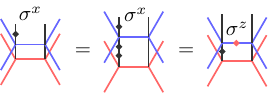}} ~,\nonumber\\
    |S_i^y\rangle&=& ~\adjustbox{valign=c}{\includegraphics{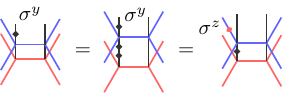}} ~,\\
    |S_i^z\rangle&=& ~ \adjustbox{valign=c}{\includegraphics{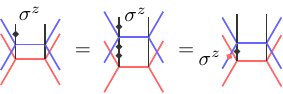}} ~,\nonumber
\end{eqnarray}
and those for the spin-dimer excited states are
\begin{eqnarray}
\label{eq:ramanEs}
    |\mathcal{K}^x_i\rangle&=&~\adjustbox{valign=c}{\includegraphics{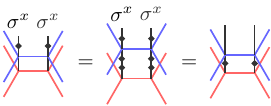}}~,\nonumber\\ 
    |\mathcal{K}^y_i\rangle&=&~\adjustbox{valign=c}{\includegraphics{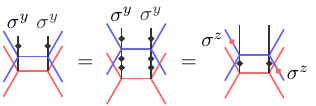}}~,\\ 
    |\mathcal{K}^z_i\rangle&=&~\adjustbox{valign=c}{\includegraphics{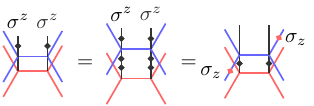}}~.\nonumber 
\end{eqnarray}
From the eigen equations in Eq.(\ref{eq:gz}) for the loop operator, we notice that each $\sigma_z$ on the virtual bond of  $\hat{Q}_{\rm LG}$ (the blue layer in Eqs.(\ref{eq:spinEs}) and (\ref{eq:ramanEs})) introduces two-visons excitations associated with the bond. Thus we are ready to show that the spin and spin-dimer excited states are the eigen-states of the vortex operator
\begin{eqnarray}
    W_p|E_i\rangle=\left\{ \begin{aligned}+|E_i\rangle, & \quad p\notin i\\~\\
    -|E_i\rangle, & \quad p\in i \end{aligned}\right. ,    
\end{eqnarray}
where $p\in i$ represents the vison excitation for the vortex operator $W_p$. The spin-excited states $|S_i^\alpha\rangle$ introduce two visons associated with the bond along $\alpha$-direction joint with the $i$-site. While the spin-dimer excited state $|\mathcal{K}^x_i\rangle$ does not induce any vortex in the vicinity of the impurity tensor, $|\mathcal{K}^y_i\rangle$ and $|\mathcal{K}^z_i\rangle$ harbor four visons.

\bibliography{draft} 
\end{document}